\documentclass[aps,prd,preprint,superscriptaddress,showpacs,preprintnumbers,amsmath,amssymb]{revtex4-1}
\usepackage{lipsum}
\usepackage{silence}
 \usepackage{booktabs}
\usepackage{graphicx}
\usepackage{epstopdf}
\usepackage{dcolumn}
\usepackage{float}
\usepackage[normalem]{ulem}
\usepackage{color}
\usepackage{dcolumn}
 \usepackage{multirow}
\usepackage{subfigure}
\usepackage{cancel}
\usepackage{hyperref}
\usepackage{url}
\usepackage{array}
\newcolumntype{L}{>{\centering\arraybackslash}m{2.7cm}}
\newcommand{\be}{\begin{equation}}
\newcommand{\ee}{\end{equation}}
\newcommand{\bea}{\begin{eqnarray}}
\newcommand{\eea}{\end{eqnarray}}

\def\lsim{\mathrel{\rlap{\lower4pt\hbox{\hskip1pt$\sim$}}\raise1pt\hbox{$<$}}}
\def\gsim{\mathrel{\rlap{\lower4pt\hbox{\hskip1pt$\sim$}}\raise1pt\hbox{$>$}}}

\begin{document}

\title{Investigating the dynamics of COVID-19 pandemic in India under lockdown}

\author{Chintamani Pai}
\email{chintupai@gmail.com}
\affiliation{Department of Physics, University of Mumbai, Vidyanagari, Mumbai, India-400098.}
\affiliation{Space Geeks, Mumbai, India.}

\author{Ankush Bhaskar}
\email{ankushbhaskar@gmail.com}
\affiliation{NASA/Goddard Space Flight Center, Greenbelt, MD, USA, 20770.}
\affiliation{The Catholic University of America, 620 Michigan Ave NE, Washington, DC 20064.}

\author{Vaibhav Rawoot}
\email{vsrawoot@mum.amity.edu}
\affiliation{Amity University Mumbai, Panvel, Maharashtra, India, 410206.}

\date{\today}
\begin{abstract}
In this paper, we investigate the ongoing dynamics of COVID-19 in India after its emergence in Wuhan, China in December 2019. We discuss the effect of nationwide lockdown implemented in  India on March 25, 2020 to prevent the spread of COVID-19. Susceptible-Exposed-Infectious-Recovered (SEIR) model is used to forecast active COVID-19 cases in India considering the effect of nationwide lockdown and possible inflation in the active cases after its removal on May 3, 2020. Our model predicts that with the ongoing lockdown, the peak of active infected cases around 43,000 will occur in the mid of May, 2020. We also predict a 7$\%$ to 21$\%$ increase in the peak value of active infected cases for a variety of hypothetical scenarios reflecting a relative relaxation in the control strategies implemented by the government in the post-lockdown period. For India, it is an important decision to come up with a non-pharmaceutical control strategy such as nationwide lockdown for 40 days to prolong the higher phases of COVID-19 and to avoid severe load on its public health-care system. As the ongoing COVID-19 outbreak remains a global threat, it is a challenge for all the countries to come up with effective public health and administrative strategies to battle against COVID-19 and sustain their economies.\end{abstract}

\maketitle

\section{\label{intro}Introduction}
Severe acute respiratory syndrome coronavirus 2 (SARS-CoV-2), a novel\linebreak coronavirus/COVID-19 outbreak was declared as a global public health emergency by the World Health Organization (WHO) on January 30, 2020 and later as pandemic~\cite{world2020time}. The virus seemed to have emerged in the population in Wuhan, China from the local wholesale seafood market in late December 2019 as per the information available in public domain~\cite{world2020report}. The rising cases in China resulted in impinging several preventive and mitigation strategies by Chinese government to control the outbreak. By January 23, all rail, road, and air transports from Wuhan city were closed and quarantine was extended to other cities in the province~\cite{mackenzie2020covid}.

\begin{figure}[h!]
\vspace*{-1cm}
\begin{center}
\includegraphics[width=\linewidth]{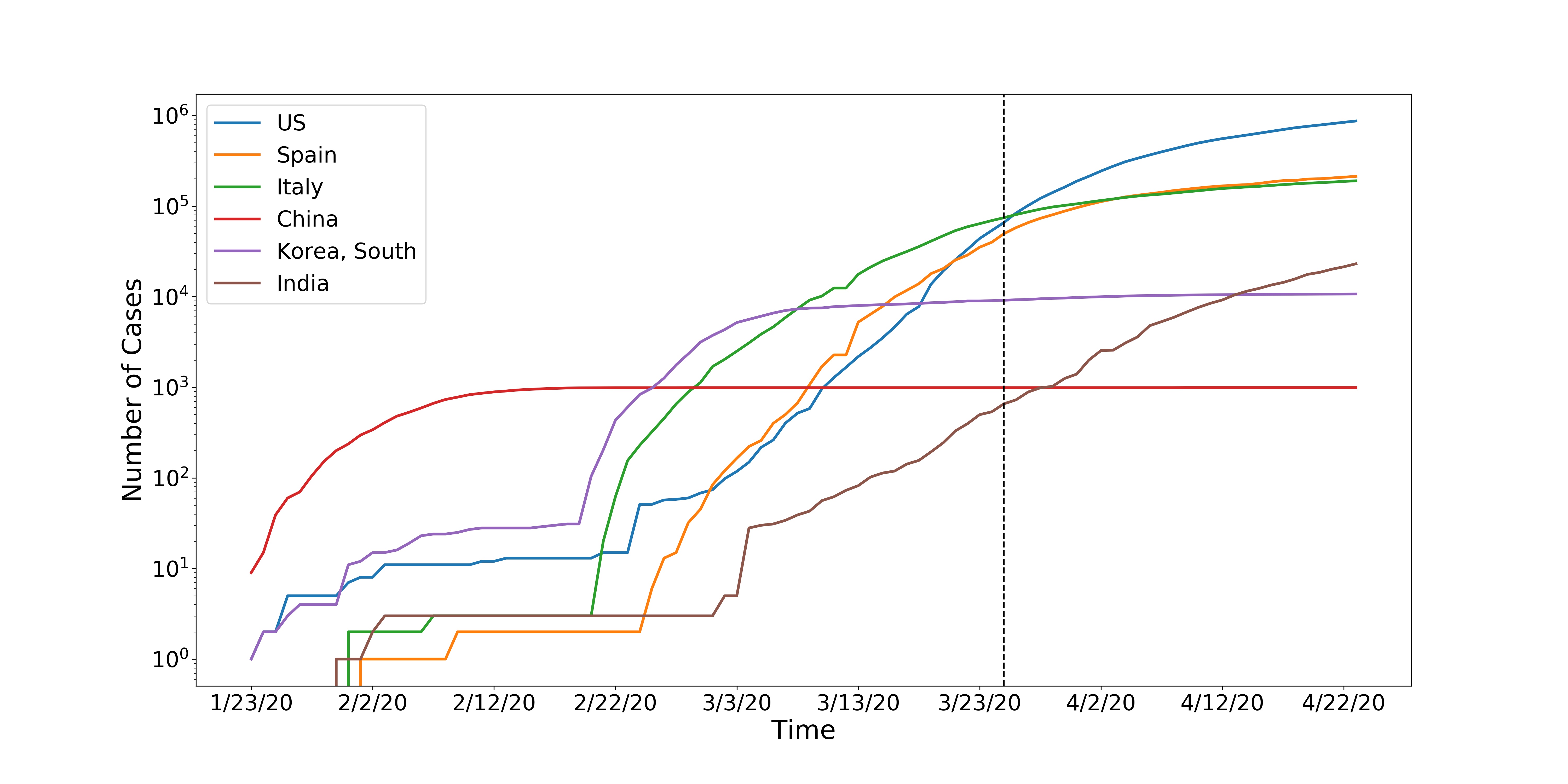}
\vspace*{-1.5cm}
\caption{Cumulative reported cases of COVID-19 for selected countries. The vertical line represents the  start of lockdown in India. China and South Korea show the flattening of cases, whereas, other countries still show increasing numbers.}
\label{fig1}
\end{center}
\end{figure}

Subsequently, rapidly rising global cases caused national emergencies in many countries putting their public health and administration departments at challenging situations. Figure~\ref{fig1} shows total cumulative confirmed cases for selected countries till date. As seen from figure~\ref{fig1}, USA, Spain and Italy are the most affected countries at present. Several other countries are facing a rapidly rising number of cases. The number of cases in China and South Korea have significantly reduced and the curve is flattened. The number of cases in India seems to be steadily growing amid nationwide lockdown. With the fast spreading of COVID-19, the global numbers have reached almost 3 million which is concerning national and international agencies. Researchers across the globe are working on developing a vaccine and antiviral drugs to fight Covid-19. UK have recently started a vaccine trial on humans~\cite{cnn2020}. Until a vaccine or anti-viral drug becomes available, suppression strategies involving public lockdown and social distancing are necessary to control the transmission though it has negative socio-economic impact. 

Strategies implemented in various countries involved a mixed combination of mitigation and suppression strategies to control the epidemic~\cite{ferguson2020impact}.  Strategies such as social distancing, closure of public and commercial activities except essential services, home quarantine for suspected cases, contact tracing and isolation of infected individuals were implemented by the government agencies across the globe. 

\begin{figure}[h!]
\vspace*{-1cm}
\begin{center}
\includegraphics[width=\linewidth]{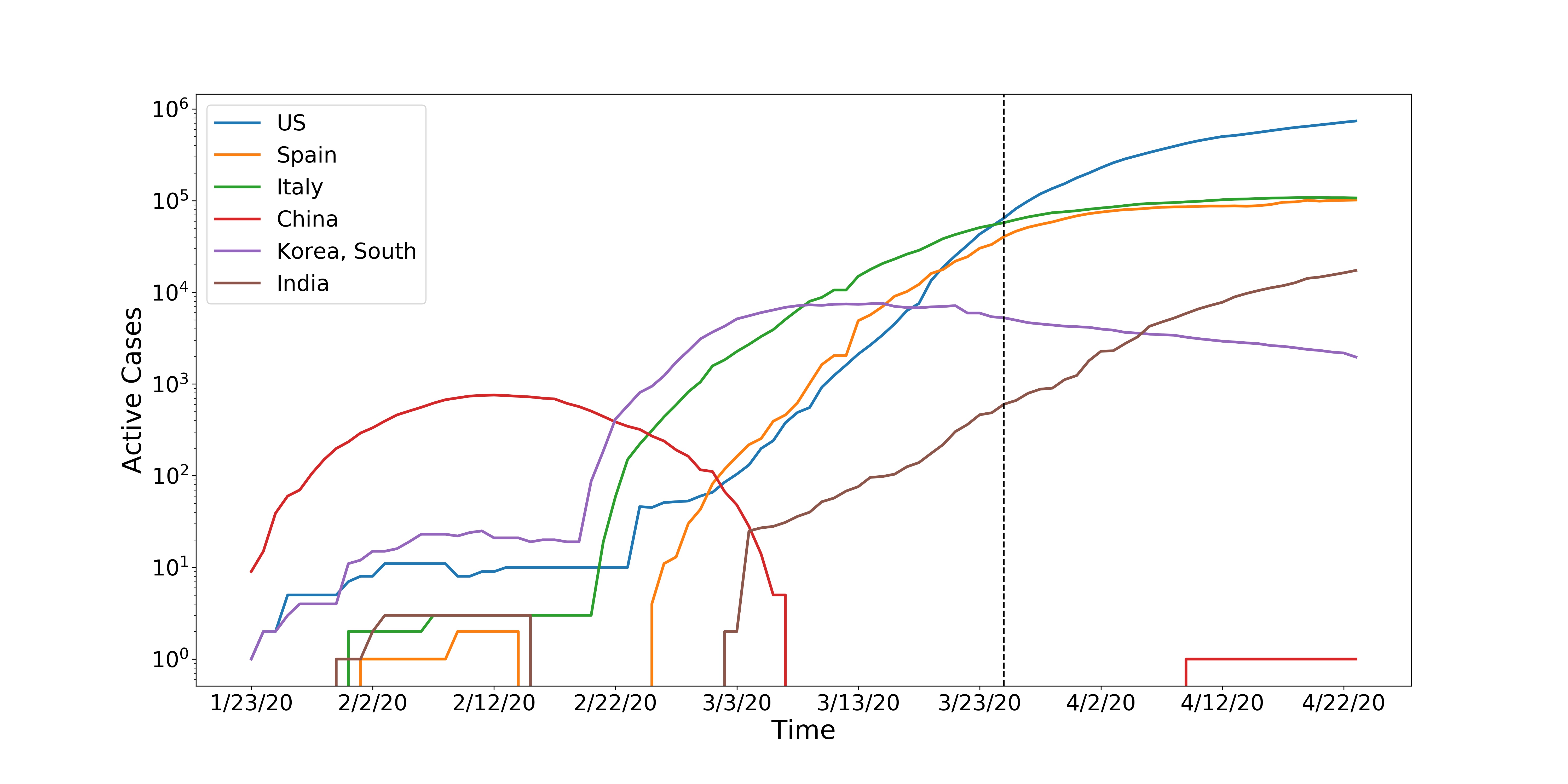}
\vspace*{-1.5cm}
\caption{Active infected cases of COVID-19 for selected countries. The vertical line represents the start of lockdown  in India. China and South Korea show the decrease in active cases, whereas, other countries still show increasing numbers of active cases.}
\label{fig2}
\end{center}
\end{figure}

Active cases on a given day is a very crucial number, which can be used to mitigate the load on available medical facilities in a country and refine effective  public health and administrative policy. Fig~\ref{fig2} shows active cases for selected countries, South Korea and China have successfully been able to  have control on this number of new cases but in whereas the number of cases  in other countries including India this number is still growing. 

For countries like India with its wide geography, dense population, non-uniform public health infrastructure across its states poses a bigger challenge to handle COVID-19 outbreak. Gauging these challenges in advance, Indian government took a smart and rapid decision going for nationwide lockdown which is considered as the globally most stringent measure to prolong the country's entry into phase 3; a community transmission mode~\cite{eco2020}. This seems to work quite well for India controlling its volume of infection compared to enormous volumes of infected cases reported for countries like Italy, Spain and USA. However, India still has some challenges standing ahead like presence of large number of asymptomatic individuals predicted by its national medical research council, increase in number of tests to be done per day and per state, reliable antibody diagnostic kits for rapid testing of suspected individuals infected with COVID-19, requirements of critical health care considering disparity in its health infrastructure and coordination among various central and state government agencies. Indian Railways have come up with an innovative approach by converting train coaches into health care facilities if a situation arises with large volumes of infected cases\cite{rail}. Health ministry in India has launched a mobile application known as ‘Aarogya Setu’ (Health Bridge) to track individual health status and currently putting efforts to increase its users~\cite{setuapp}. India’s national medical council has recently launched a pool testing method for RT-PCR diagnosis to narrow down local areas affected with COVID-19 cases which help in identifying infected individuals for their immediate isolation~\cite{icmr:report1}. 

India announced US $\$$2.1 billion aid for the health sector in India to fight the COVID-19 situation. The Department of Science and Technology, Government of India has come up with schemes to promote research and inventions in research institutes and start-ups working in various domains which will help in controlling the outbreak as the country is undergoing the lockdown. India’s Council of Scientific and Industrial Research (CSIR) with its 38 labs across the country geared up to mitigate the COVID-19 situation with their five pronged approach~\cite{csir:covid1}. Total 
5,79,957 tests (as of April 25, 2020) of samples were carried out in India as per the report of Indian Council of Medical Research (ICMR)~\cite{icmr:report2}. ICMR launched nationwide COVID-19 laboratories to test the samples including few private labs with appropriate safety and health protocols for testing the samples~\cite{icmr:testing}. Recently India also introduced blood plasma therapy for COVID-19 which aims at using the blood plasma of recovered individuals who have gained immunity against COVID-19 and transferring it to infected individuals~\cite{pib:plasma}. Indian Space Research Organization (ISRO) has also developed COVID-19 INDIA map showing geo-spatial data of COVID-19 cases~\cite{isro}. Defence Research and Development Organisation (DRDO), India  has focussed on making critical care medical requirements and transferred some of its technologies to private companies for medical supply during COVID-19 situation~\cite{drdo}.   

At present, the overall volume of infected numbers in India is definitely less compared to other countries due to timely announcement of lockdown and considerable efforts by Indian authorities to impose stringent lockdown. Yet there are challenges lying ahead while the country is undergoing lockdown and these challenges may get further amplified after the lockdown is removed (fully or partially) on May 3, 2020. 

In this paper, we investigate the current lockdown situation by setting up Susceptible-Exposed-Infectious-Recovered (SEIR) epidemiological model applied to ongoing COVID-19 spread in India to forecast the growth of active infected cases. Based on this modelling, we provide a macro perspective by gauging the situation if lockdown is removed after May 3, 2020. Based on our results we also emphasize on strategies to be taken in order to curb the spread of COVID-19. In the next section, we do a brief review of studies carried out describing the growth of COVID-19 spread in India.  

\section{Review of COVID-19 prediction studies for India}
COVID-19 cases in India have been rising rapidly and can create a severe impact on the country's healthcare sector and the economy. India reported the first case of COVID-19 on January 30, 2020 in Kerala, a state located in the southern part of the country and the total number of cases crossed 20,000 on April 21, 2020. Multiple studies involving epidemic models and other mathematical modelling approaches have been published and reported in the media. These studies include research publications by researchers, financial/data consulting firms, independent organizations working in the field of epidemic control. In this section, we briefly review these studies describing possible scenarios for the COVID-19 outbreak in India.  
 
The Center for Disease Dynamics, Economics $\&$ Policy (CDDEP), a public health research institute carrying out independent, multidisciplinary research in the public health sector to assist in the decision making process for public health policies. CDDEP has given a prediction for COVID-19 outbreak in India under the model name IndiaSIM [19]. IndiaSIM model predicted a number of infections for different lockdown strategies in India. Their model suggests the peak number of infections will be around 1.5 million and 1 million for no interventions and hard lockdown respectively. It is mentioned that the peak will occur at the end of June 2020 with hard lockdown and beginning of June 2020 with no interventions. The IndiaSIM model assumed that the 21-day lockdown would reduce disease transmission by 25$\%$ in moderate lockdown or 44$\%$ in hard lockdown conditions. 
Protiviti, a global consulting firm has predicted numbers of COVID-19 cases crossing 75,000 by May 22, 2020~\cite{protiviti}. Their study made use of a combination of models such as percentage model, time series model and SEIR model. Data Driven Innovation Labs from Singapore University of Technology and Design has also come up with a dashboard called ‘When COVID-19 will end’ providing predictive monitoring of COVID-19 cases for multiple countries including India. Their SEIR based data regressive model predicts that COVID-19 in India will end by June first week~\cite{sutd:results1}.  

COVID-19 India (\url{https://www.covid19india.org/}) is a data portal providing updated information on COVID-19 cases in India.  Creators of this portal have recently published their work~\cite{debashree:covid19} on the analysis of COVID-19 cases using SIR and eSIR models talking about timely decisions made by the Indian Government to control the COVID-19 growth. Paper also discusses the possible scenarios once the lockdown is removed based on different combinations of control strategies such as lockdown removal with moderate release, travel ban, social distancing etc. to predict the growth in COVID-19 cases after May 3.

SEIR and regression model based COVID-19 outbreak predictions are given by Ranjan Gupta et al. in the ref. [cite Ranjan Gupta et  al.]. The reproduction number calculated by them is 2.02. The effect of lockdown and social distancing is estimated using an age structured Susceptible Infected Recovered (SIR) model in Ref.~\cite{rajesh:agestr}. Their study has obtained the reproduction number equation to 2.108 and It is found that the most susceptible population is in the age group of 0-50 year whereas the mortality rate will be highest in the age group 50-80 year. Their study also explores the different lockdown strategies predicting increase in number of cases. Several other studies based on epidemic models are also presented to give predictions of COVID-19 spread in India~\cite{ranjan:paper1,kaur:paper1,gupta:paper1}.    

We explore the macroscopic perspective based on the SEIR model to understand the scenario of COVID-19 in India, in the background keeping this review of studies for predicting COVID-19 India dynamics. 

\section{Description of mathematical model to predict the dynamics of COVID-19}
We start with the SEIR model to investigate the spread of COVID-19 in India. Epidemiological models like SEIR provide a useful quantitative approach to understand the macroscopic picture of the disease spreading. The underlying philosophy is to divide the population under consideration in compartments of individuals of types susceptible S, exposed E, infected I and recovered R. An individual in the model does the transition from one compartment to another based on underlying parameters which govern the dynamics of disease spread in the population. In our present work, we avoid to include any heterogeneities in the context of Indian population and COVID-19 in our model and make an attempt  to understand the dynamics with the basic SEIR model described in this section. 

The SEIR model contains an additional parameter, the latent period. During the process of infection spreading, there is an incubation period during which an individual host is having a pathogen which is reproducing itself inside the host at a faster rate but yet to reach an active transmission mode to infect other susceptible hosts in the surroundings. Hence, a new compartment of exposed individuals E is introduced, which contains individuals who are infected but yet to become infectious. The SEIR model is an extensively used epidemiological model based on the work of Kermack and McKendrick~\cite{kermack1927}.  The SEIR model reflects the dynamics of flow of individuals among four compartmental states; S, E, I and R. The model is described below using the coupled ordinary differential equations. 
\begin{eqnarray}
 \frac{dS}{dt} &=& -\beta S I	\\
 \frac{dE}{dt} &=& \beta S I - \sigma E \\
 \frac{dI}{dt} &=& \sigma E - \gamma I \\
 \frac{dR}{dt} &=& \gamma I
\end{eqnarray}
In the above equations, $\beta$, $\sigma$ and $\gamma$ denote the rates of transmission, incubation and recovery respectively. We also assume that the susceptible population at any time during the dynamics follows the condition given below,
\begin{equation}
 N = S + E + I + R
\end{equation}
where, N is the total population at the start of disease spread. In present work we neglect general the birth rate and death rate of population. In terms of the parameters already described above, the Reproduction Number $R_0$ is defined below  as
\begin{equation}
 R_0 = \frac{\beta}{\gamma}
\end{equation}
Reproduction Number $R_0$ has a central importance in Epidemiology providing transmission potential of infectious disease and it tells about expected number of new infections transmitted by single infected individuals entered in susceptible populations. The transmission potential of infectious disease may change over time and thus, $R_0$ may also change over time. There are several ways to estimate the variation of $R_0$ with time and in this paper we restrict ourselves to not get into the methods of $R_0$ estimation. Instead, we optimize the parameters of the SEIR model described above to generate plots for active infected cases based on the observed data of COVID-19 cases. Continuing with this approach, we also analyze the effects of lockdown to project the growth of active infected cases during the lockdown and after it is removed. Based on the SEIR model, we discuss the effectiveness of control strategies.  

The value of $R_0$ for COVID-19 is estimated to be between 2 and 3~\cite{callway:ro}. Another approach of computing $R_0$ involves the processing discrete time series data for active infected cases, recovered cases and deaths. For advanced analysis, the SEIR model can be further modified to estimate the effect of various conditions such as quarantine, meta populations and presence of asymptomatic cases.       
\begin{figure}[h!]
\begin{center}
\includegraphics[width=\linewidth]{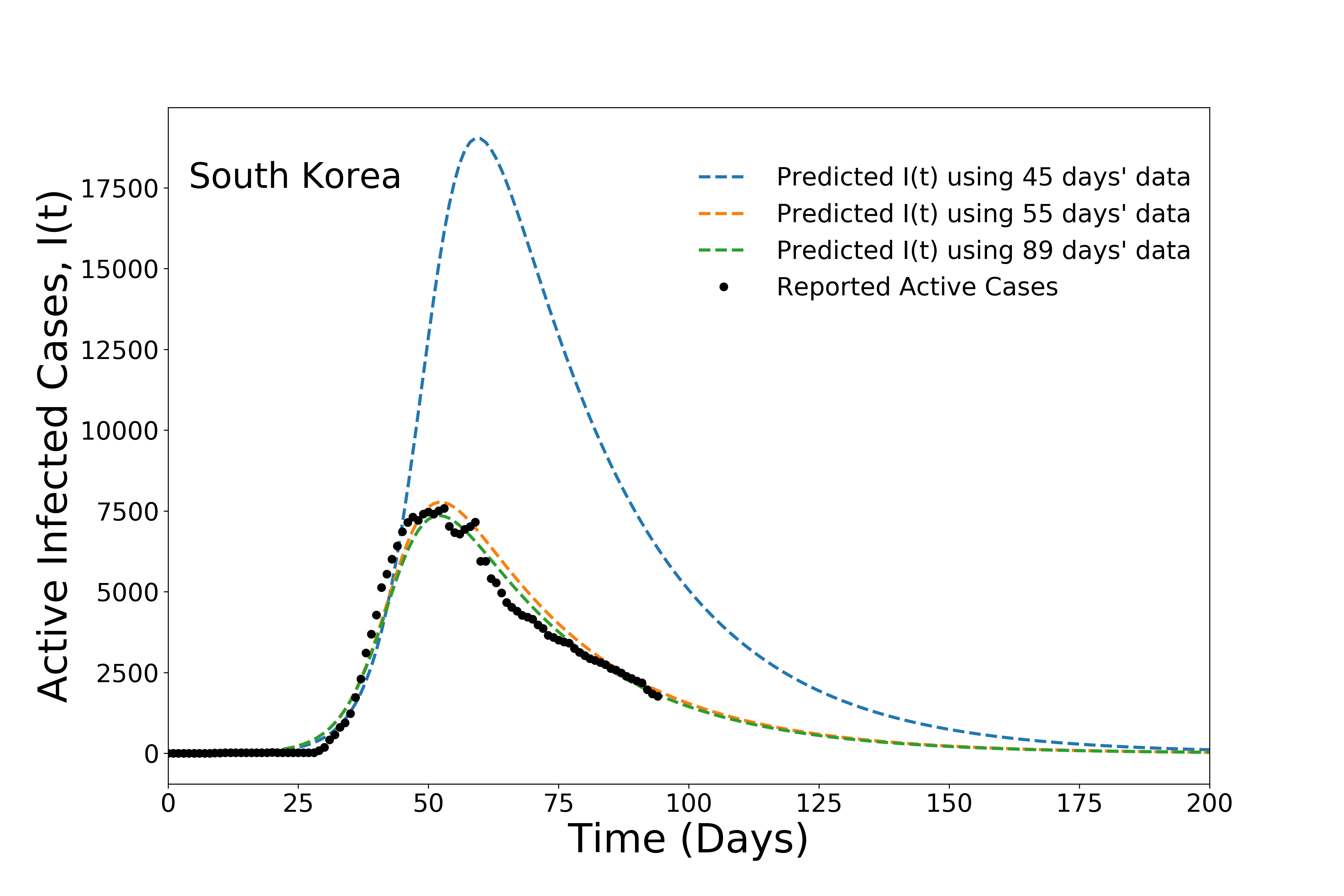}
\vspace*{-1.5cm}
\caption{Predicted active infected cases (dashed lines) of COVID-19 for South Korea for varying data size are shown.   The first day is on January 22, 2020.  Parameters correspond to recovery period and incubation period are $\gamma$=0.0385 and a=0.2 respectively}
\label{fig3}
\end{center}
\end{figure}

\begin{figure}[h!]
\begin{center}
\includegraphics[width=\linewidth]{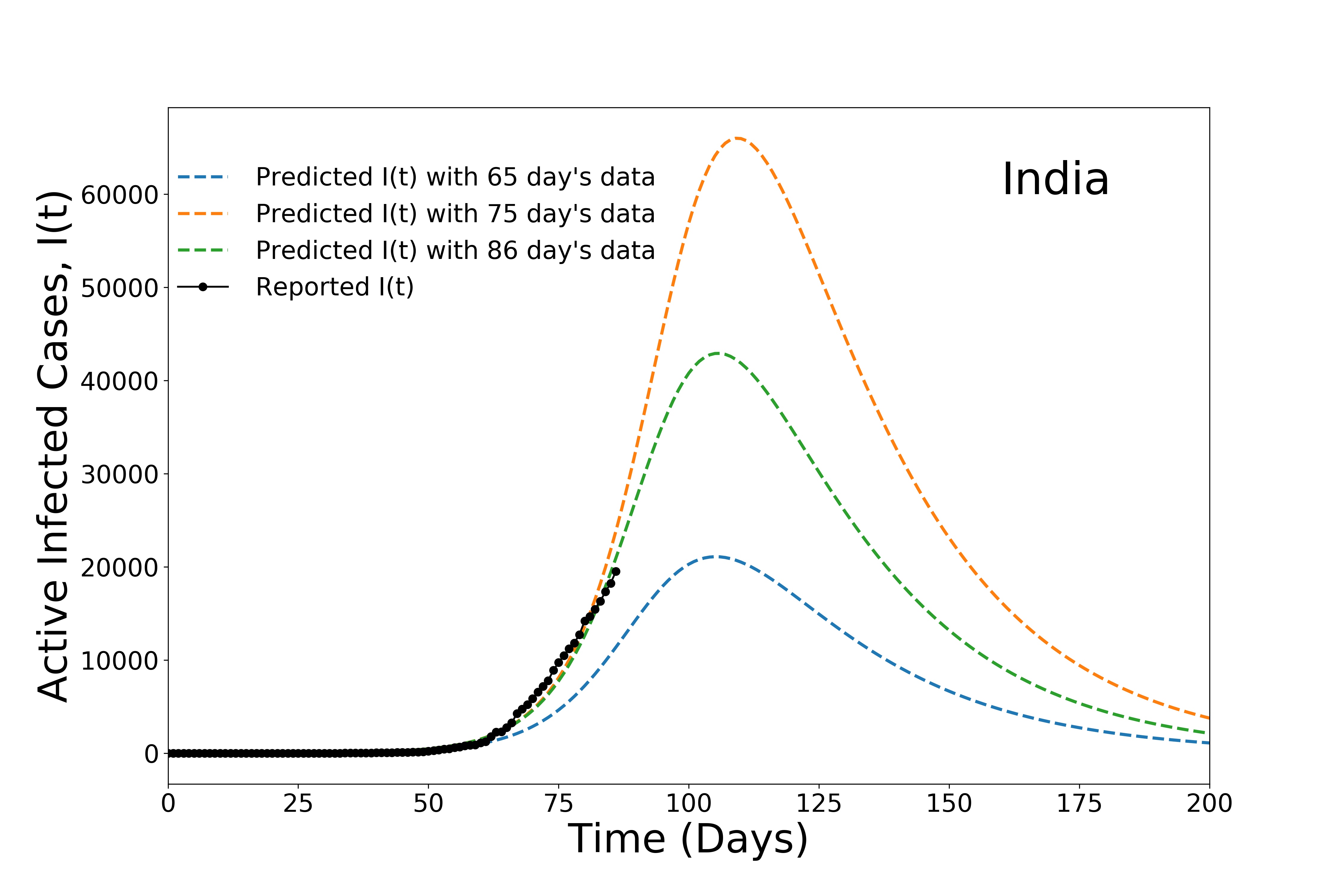}
\vspace*{-1.5cm}
\caption{Predicted active infected cases of COVID-19 for India considering varying length of data (dashed curves).  Reported active cases are shown in black solid circles. The first day is on January 30, 2020.  Parameters correspond to recovery period and incubation period are $\gamma$=0.0385 and a=0.2 respectively.}
\label{fig4}
\end{center}
\end{figure}

\begin{figure}[h!]
\begin{center}
\includegraphics[width=\linewidth]{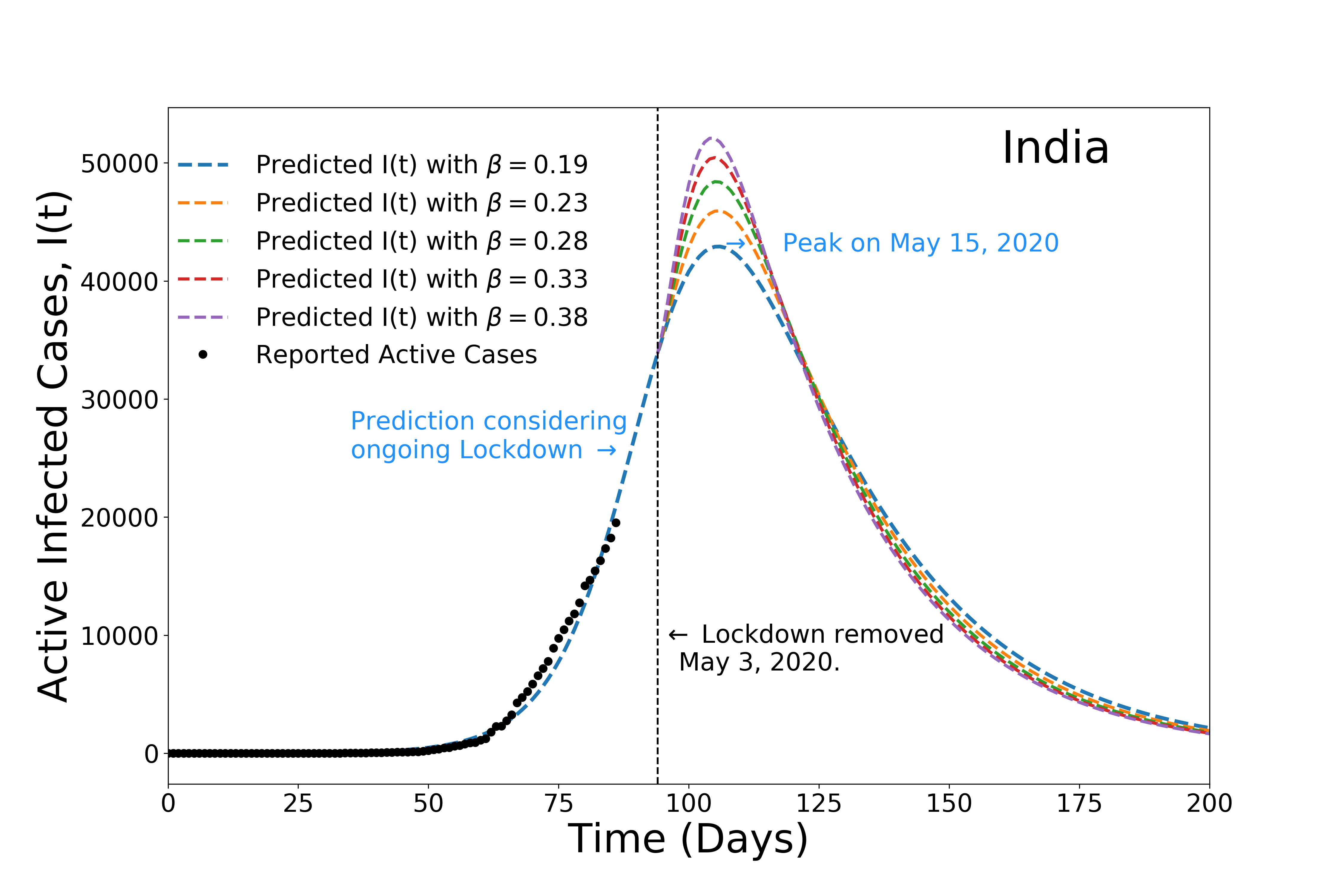}
\vspace*{-1.5cm}
\caption{Predicted active infected cases of COVID-19 for India undergoing lockdown (blue dashed curve,  Beta=0.19) and possible scenarios after the lockdown removal on May 3, 2020 (For Beta=0.23, Beta=0.28,  Beta=0.33,  Beta=0.38).  Reported active cases are shown in black dots. The first reported case in India was on January 30, 2020.  Parameters correspond to recovery period and incubation period are $\gamma$= 0.0385 and $\sigma$ =0.2 respectively.}
\label{fig5}
\end{center}
\end{figure}
\begin{table}[t]
\begin{tabular}{|L||L|L|L|L|L|}
\hline
  & Projected Based on data during lockdown &\multicolumn{4}{|c|}{Projected if the lockdown is removed on
 May 3, 2020} \\
 \hline \hline
$\beta$ & 0.19 & 0.23 \linebreak(25$\%$ increase) & 0.28 \linebreak(50$\%$ increase) & 0.33 \linebreak(75$\%$ increase) & 0.38 \linebreak(100$\%$ increase)\\
 \hline
Peak infected active cases & 42918 & 45939  & 48411 & 50447 & 52086\\
 \hline
$\%$  increase in  active infected cases  & NA & 7 &  13  & 18 & 21.3\\
 \hline
Peak date of active infected cases & May 15, 2020 & May 15, 2020 & May 14, 2020 & May 14, 2020 & May 13, 2020\\
 \hline
\end{tabular} 
\caption{Comparison of infected cases after Lockdown  removal with respect to projected cases based on ongoing lockdown.}
\end{table}

\section{Data Analysis, Model Fitting and Results}
The data of confirmed, recovered and death cases due to  COVID-19  spread  is obtained from  \url{https://github.com/CSSEGISandData/COVID-19} repository. This  data repository for the 2019 Novel Coronavirus Visual Dashboard is operated by the Johns Hopkins University Center for Systems Science and Engineering (JHU CSSE). Also, it is supported by ESRI Living Atlas Team and the Johns Hopkins University Applied Physics Lab (JHU APL). The active infected cases are estimated by subtracting recovered and death cases from the confirmed cases. Python open source programming is used for analysis and fitting the data.

The SEIR model is fitted with the data to predict the possible active cases in India. In the case of South Korea, the cumulative curve of COVID-19 cases shows the complete dynamics with initial exponential growth and then later flattening of the curve.  Therefore, we decided to use the COVID-19 data of South Korea to test the efficiency and feasibility of our SEIR model and fitting procedure. For fitting our model to South Korea, certain parameters in the SEIR model like transmission rate, $\beta$ were optimized during the fitting procedure. We have used the incubation period of 2 days ($\gamma$=0.2) and recovery time as 26 days ($\sigma$ = 0.0385) for our fitting~\cite{fitpara}. 

For optimization of our fit, we have minimized  chi-square for varying values of  $\beta$ and initial susceptible population (N).  This helped us to determine pairs of N and $\beta$ which gave the best fit to the data for active infected cases.The chi-square is given by the following equation,
\begin{equation}
 \chi_c^2 = \sum \frac{O_i - E_i}{E_i}
\end{equation}
where $O_i$ is observation and $E_i$ is modeled/expected value for a given instant.  The fitting of a model is highly dependent on the length of observed data taken into consideration. Fitting the model for small length of the data especially during the onset of exponential growth, the number of cases is low compared to expected numbers during the complete dynamics. Hence, fitting may not be very accurate for a smaller length of the data for any forecasting. 
Figure~\ref{fig3}, shows fitted (dashed lines) model and observed (solid dots) active cases of South Korea for varying data length. Note that, predicted values by the model for observed data till 55 and 89 days converge to similar values and very close to the reported cases, ensuring confidence in our optimization procedure of the parameters.

After testing our model on South Korea, we then extended it to India. As COVID-19 cases started emerging in India a bit later in time compared to its onset in other countries, the COVID-19 outbreak  in India is in the growth phase. India has implemented hard lockdown very early which seems to help in controlling the transmission. To give the projection of infected active cases based on current lockdown condition and to investigate if the lockdown is removed on May 3, 2020 how the curve will evolve, we have fitted the SEIR model with chi-square optimization.  First, we fitted the model for varying size of data. Figure~\ref{fig4} shows the predicted curves from this analysis. The variation in predicted curves due to varying size of data is evident. However, considering current data of 86 days from its onset in India, the predicted curve(dashed green line) shows a nice fit to the reported cases. Assuming the projected trends of 75 and 86 days are more reliable, we observe that peak active infected cases could reach around ~ 42000-70000 by mid May, 2020 in India. 

Moreover, we evaluated the impact of the possible lockdown removal on May 3, 2020 using our model. The $\beta$ value for the ongoing lockdown period is 0.19, we assumed that the $\beta$ could increase after lockdown removal. The $\beta$ value will change as control measures are removed either fully or partially~\cite{fang2020tdot}.  It is difficult to quantify the exact value of beta after lockdown period, therefore, we project possible scenarios for $25\%$  ($\beta$=0.23), $50\%$  ($\beta$=0.28), $75\%$  ($\beta$=0.33), and 100$\%$  ($\beta$=0.38) increase in the current $\beta$ value (0.19). Table 1 summarizes the projected active infected cases if  lockdown is continued and different scenarios for post-lockdown situations. Figure~\ref{fig5} depicts  the curves for different $\beta$ values  forecasting the scenario after lockdown removal in comparison to the predicted curve for active infected cases (dashed blue line)  based on  current data.

\section{Discussions and Conclusion}
In our present work, we analyze the observed data of COVID-19 cases in India using the SEIR model. To start with, we used data of COVID-19 from South Korea showing all the phases of active infected cases over time to validate our model. Then, we used the chi-square minimization process to obtain best fitted parameters for South Korea and India. This enabled us to forecast  the peak of active infected cases for India. It is found that with respect to the ongoing lockdown situation in India, the peak of active infected cases will occur in Mid of May, 2020 with a maximum number of  ~ 43,000. Our study shows that the infected cases will tend to cease by the mid of August, 2020 if lockdown is continued. Interestingly, our predicted peak time is very much consistent with the predictions made by Protiviti, a global consulting firm. In our model, we consider different values of beta (other than for the lockdown condition; $\beta$ = 0.19) to provide insight on the relative strengths of control measures taken by the government. Lower the value of beta, indicates a relatively better strategy to control the spread. The removal of lockdown on May 3, 2020 may lead to $\sim$ 21 $\%$ increase (corresponding value of transmission rate, $\beta$ = 0.38) in the peak value of active infected cases as compared to the peak value predicted for ongoing lockdown. This increase is due to the exposed individuals or asymptomatic individuals who may actively transmit in post-lockdown period.  Therefore, it is highly important to identify such exposed and asymptomatic individuals in post-lockdown period. Increasing the number of tests for early identification of COVID-19 cases during the lockdown period will certainly help in understanding the real picture of infected volume. Mobile applications or web GIS applications integrated with GPS which are capable of tracking regular geo-spatial movements and health status of individuals can provide useful data~\cite{spacegeeks}. If an individual is found infected, such applications will help to trace the possible spread of disease in the post-lockdown period.  

India’s bold and timely decision going  for a nationwide lockdown of 40 days has certainly helped the country in controlling the infection volume at the cost of adverse socio-economic impact.  Whether full or partial lockdown remains after May 3, severely affected zones (states/districts/villages/cities) in the country will significantly contribute to the overall growth of COVID-19 cases in India. Lockdown removal strategies have to be smartly designed which do not allow estimated infected volumes centered around these zones to grow. Model based forecasts of peak values will certainly help decision makers, especially public health and administrative departments of the state and central governments to assess their preparedness, control the pandemic and repair the socio-economic damage caused by the pandemic. Lockdown period is a period not only to control the pandemic but a time bought to re-design the future national and international policies to ensure the sustainable growth of the economy.  


\bibliography{ref.bib}
\bibliographystyle{unsrt}
\end{document}